# Interpretation of Stark broadening measurements on a spatially integrated plasma spectral line

J. Thouin, M. Benmouffok, P. Freton, J.-J. Gonzalez

LAPLACE, Université de Toulouse, UMR 5213 CNRS, INPT, UPS, Toulouse, France

Email : thouin@laplace.univ-tlse.fr

# 1 Abstract:

+

In thermal plasma spectroscopy, Stark broadening measurement of hydrogen spectral lines is considered to be a good and reliable measurement for electron density. Unlike intensity based measurements, Stark broadening measurements can pose a problem of interpretation when the light collected is the result of a spatial integration. Indeed, when assuming no self-absorption of the emission lines, intensities simply add up but broadenings do not. In order to better understand the results of Stark broadening measurements on our thermal plasma which has an unneglectable thickness, a Python code has been developed based on local thermodynamic equilibrium (LTE) assumption and calculated plasma composition and properties. This code generates a simulated pseudo experimental (PE) H$_\alpha$ spectral line resulting from an integration over the plasma thickness in a selected direction for a given temperature profile. The electron density was obtained using the Stark broadening of the PE spectral line for different temperature profiles. It resulted that this measurement is governed by the maximum electron density profile up until the temperature maximum exceeds that of the maximum electron density. The electron density obtained by broadening measurement is 70% to 80% of the maximum electron density.

# 2 Introduction:

For the experimental characterisation of an electric arc, emission spectroscopy is the diagnostic method of choice for the measurement of the plasma temperature or species densities. Among the different spectroscopic diagnostic methods, those based on the measurement of the Stark broadening of a spectral line are widely used [1–6].

In order to perform a spatially resolved spectroscopic measurement within the plasma there are several methods. When an assumption of plasma axisymmetry can be made, methods based on Abel inversion can be used [6, 7]. When this assumption is no longer valid and the optical system allows simultaneous acquisitions in different directions, tomographic reconstruction methods can be considered [8].

However, when none of these methods can be applied, because the plasma is not axisymmetric, the time constraints are too great, the intensity of the discharge is not sufficient to allow the selection of light rays or because the experimental setup does not allow a complex optical setup, it is common to collect the light emitted by the plasma using a focusing optical setup or selecting a direction with pinholes [1, 3, 5, 9]. In this case the light is collected over the entire thickness of the plasma and the broadening measurement is made on the spectral line resulting from the integration over the thickness of the plasma. For inhomogeneous plasma and depending on the measurement performed, it can be complex if even possible to interpret the measured quantity.

The presented study was performed in order to interpret experimental results of Stark broadening measurements on $H_\alpha$ spectral line emitted by a water thermal plasma. This plasma, considered at local thermodynamic equilibrium (LTE), is generated by a ten milliseconds electric arc in a water tank which vaporizes the water. The light is collected in a direction selected using a couple of pinholes. We are interested in the theoretical determination of the electron density by measurement of the broadening of a spectral line. We are interested in the hydrogen spectral line $H_\alpha$ for which we consider the Stark effect as the dominant source of broadening in the presence of a high electron density [2].

In order to study the relevance of the broadening measurement to obtain the electron density, we will first define the context of our study. Then, we will present the parameters that govern the profile of a spectral line: (1) the emissivity of the transition associated with the $H_\alpha$ spectral line which will be calculated from the plasma composition and (2) the Stark broadening of the Hα line determined by simulation from the work of Gigosos *et al.* [10] We will then perform a parametric study using different temperature profiles for the water thermal plasma. We will focus on the electron density determined from the reconstructed Hα pseudo experimental (PE) spectral line for these different profiles assuming that the broadening is only due to Stark effect. Finally, we will conclude on the meaning of the obtained measurements.

## 3    Context of this study:

We performed experimental emission spectroscopy measurements on a water plasma. This plasma was generated between two vertical sharpened rods of tungsten (see fig. 1). In order to generate the arc, a half sine wave of current is applied through a fuse wire. The wire is vaporized by Joule effect and generates a water vapour bubble which gradually expands before collapsing. The duration of the discharge is 10ms and the sinusoidal current wave (f=50Hz) has an amplitude of about 1kA. A theoretical study of the phenomenon was carried out in our team [11] in order to understand the behaviour of the plasma. A water plasma bubble was simulated using the commercial @Fluent software based on the finite volume method. Plasma properties have been calculated for water, the theory is presented in Harry-Solo *et al.* [12]. First instants of the bubble formation are not described; the simulation begins with a conducting channel already established. Using the experimental variations of the measured voltage and current intensity, a source term is applied within a volume defined by a boundary temperature of 7kK. This arbitrary temperature defines the conducting channel. Naturally, this volume changes during the deposition of energy. The phase transition between vapour and liquid was handled using a model based on that of Lees [13]. This study gave us a homogenous value for the pressure inside the vapour bubble close to 3 bar at very early times of the expansion (t<0.5ms). Therefore, in this article the plasma will be considered to be pure water and its properties calculated for a pressure of 3 bar.

To illustrate, fig. 1 shows the observed bubble with the plasma contained in the saturated zone. This image was obtained using a Photron FASTCAM SA5 high-speed camera, the exposure time was 1/25000s and three neutral density filters were fitted on the lens with an attenuation factor respectively of 64, 32 and 16. The acquisition method is described with more details in [14]. The circle in the foreground corresponds to the shape of the observation window. A halogen lamp is used as backlighting, its filament is visible horizontally in the background. Backlighting is only necessary to improve the observation of the edge of the water vapour bubble. Two copper electrode holders support the tungsten rods of 1.6 mm diameter vertically. Between the two electrodes is placed a 0.13 mm copper wire.

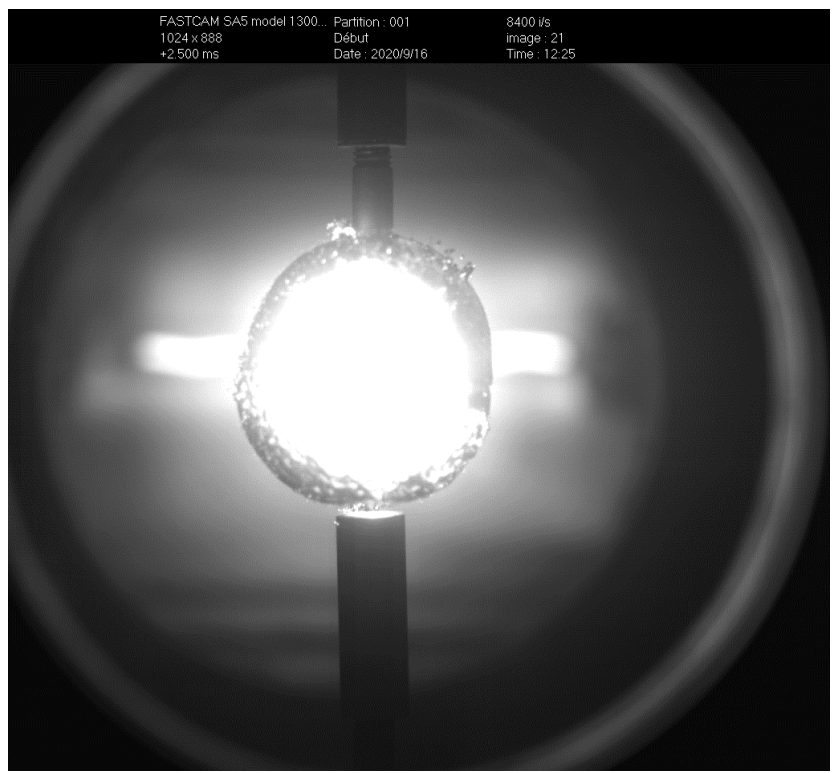

Figure 1: Image of a water vapour bubble. The image is saturated in the centre by the emission of light from the arc.

The light emitted by the plasma is collected through two irises which select a beam of light. When considering no self-absorption, the light collected is the sum of all emissions on the line of sight selected by the two pinholes, see fig. 2:

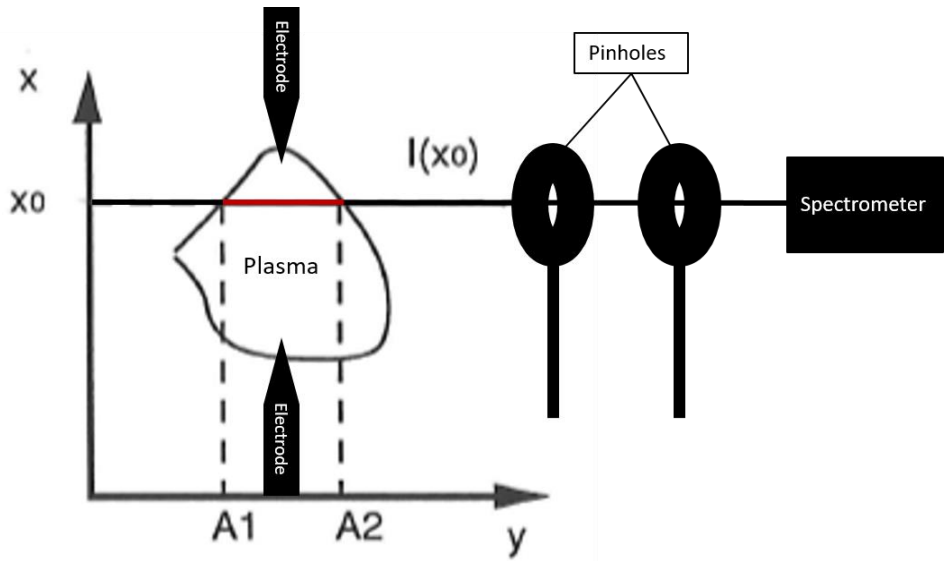

Figure 2: Plasma light selection and acquisition.

Figure 2 shows the light being collected on the line of sight I(X0). We can consider that the profile of one emission spectral line, at a given wavelength, integrated over this string of plasma is the sum of the local emissions at that same wavelength along the string. These local emission spectral lines can be characterized using the local conditions of emission. We will describe this process in sect. 4.

## 4   Spectral emission line description:

The shape of a spectral line profile can be characterized using different parameters such as its amplitude or its broadening. The area covered by the profile of the spectral line is dependent on those two previous parameters but also on the shape of the profile. These different parameters are illustrated on fig. 3. The full width half area FWHA of the spectral line is the width of the hatched area. The full width at half maximum FWHM is plotted as well.

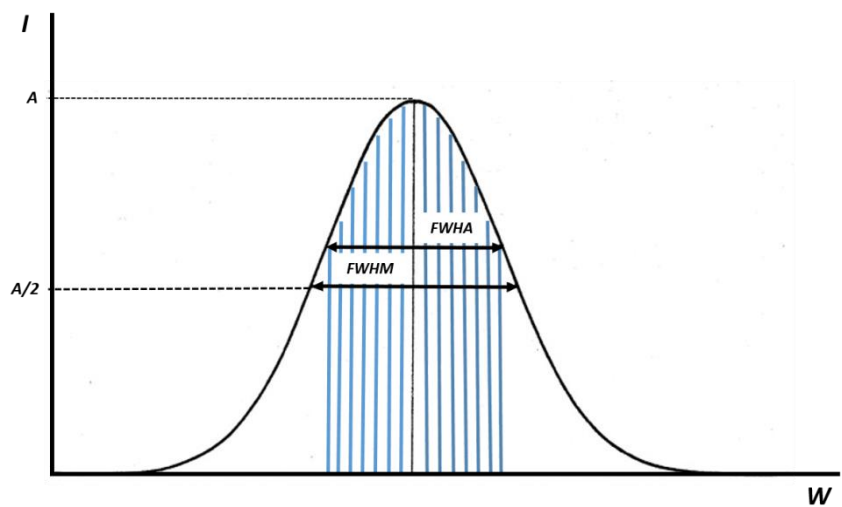

Figure 3: Intensity of a given spectral line plotted against the wavelength; Illustration of FWHM and FWHA parameters.

## 4.1 Broadening:

The physical causes of spectral line broadening in a thermal plasma are numerous, however, a natural broadening (small) as well as a broadening due to the optical system and finite resolution of the spectrometer, are always present. In addition, depending on the plasma and spectral emission line considered, Doppler, Van der Walls or Stark broadenings can be present. Depending on its cause, the broadening profile will be either Gaussian (for the optical and Doppler broadenings) or Lorentzian (for the natural broadening and collisional broadenings such as Van der Walls or Stark). The different causes of broadening of a spectral line are discussed in detail in the work of H. Griem [15]. The resulting spectral line profile is a Voigt profile, *i.e.* a convolution of a Lorentzian and a Gaussian profile. It can lean towards a Gaussian or a Lorentzian profile depending on the sources of broadening. If several sources of broadening are present, some of them depending on the emission environment and therefore the position of emission, reconstructing the profile of the integrated line can be difficult.

In this work, we are interested in the Hα spectral line for which the Stark broadening is such that all other sources of broadening are negligible [2]. The resulting profiles tend to be Lorentzian.

The broadening of Hα is provided to us directly as a function of the electron density $n_e(x,y,z)$ by the equation determined by Gigosos *et al.* [10] with the FWHA:

$$FWHA = 0.549 \left(\frac{n_e}{10^{23}}\right)^{0.67965} \tag{1}$$

A Lorentzian is defined as:

$$f(\lambda, A, \sigma) = \frac{A}{\pi} \frac{\sigma}{\sigma^2 + \lambda^2} \tag{2}$$

With:

- $\lambda$ wavelength
- $A$ amplitude factor
- $2\sigma$ FWHM
- $n_e$ electronic density

The indefinite integral is:

$$F(\lambda; A, \sigma) = \frac{A\sigma}{\pi} \tan^{-1}\left(\frac{\lambda}{\sigma}\right) \tag{3}$$

And the area $a_l$ of the profile is:

$$a_l = \int_{-\infty}^{+\infty} f(\lambda, A, \sigma) \, d\lambda = A \, \sigma \tag{4}$$

Let's assume that $\Delta_{\frac{1}{2}}$ gives the FWHA so that:

$$\frac{a_l}{2} = \frac{A\sigma}{2} = \int_{-\Delta_{\frac{1}{2}}}^{+\Delta_{\frac{1}{2}}} f(\lambda, A, \sigma) \, d\lambda \tag{5}$$

$$\frac{A\,\sigma}{\pi}\left(\tan^{-1}\left(\frac{\Delta_{\frac{1}{2}}}{\sigma}\right) - \tan^{-1}\left(\frac{-\Delta_{\frac{1}{2}}}{\sigma}\right)\right) = \frac{A\,\sigma}{2} \tag{6}$$

Which is true for: $\tan^{-1}\left(\frac{\Delta_{\frac{1}{2}}}{\sigma}\right) = \frac{\pi}{4}$ therefore $\frac{\Delta_{\frac{1}{2}}}{\sigma} = 1$ and $\Delta_{\frac{1}{2}} = \sigma$.

It should be noted that the FWHA is equal to $2\sigma$, which is also the FHWM for a Lorentzian profile.

### 4.2 Emissivity:

In order to reconstruct the integrated spectral line emitted by the plasma in one direction, the amplitude of the spectral line or its intensity (area) is necessary in addition to the broadening profile. This information is provided by the emissivity of the spectral line. The emissivity associated with an electronic transition from an energy level i to an energy level j is expressed as:

$$J_{tot} = \int_{-\infty}^{+\infty} J(\lambda) \, d\lambda = n_i(P,T) \, A_{ij} \frac{hc}{\lambda} \frac{1}{4\pi} \tag{7}$$

With:

- $J(\lambda)$ emissivity profile as a function of the wavelength
- $n_i$ population density of the emitting level i
- $A_{ij}$ Einstein coefficient for the electronic transition from level i to j
- P, T pressure and temperature
- h, c Planck's constant and speed of light in vacuum

If the plasma is assumed to be at local thermodynamic equilibrium (LTE) then the density $n_i$ of the level i is given by Boltzmann law:

$$n_i(P,T) = \frac{n(P,T)}{Z(P,T)} g_i e^{\frac{-E_i}{KT}} \tag{8}$$

With:

- $n(P,T)$ global species density: $n(T,P) = \sum_k n_k$ with k being the excitation level.
- $Z(P,T)$ partition function for the considered species

- $E_i$ energy for the level i
- $g_i$ degeneracy of the level i
- $K$ : Boltzmann constant

The emissivity can be expressed as:

$$J_{tot}(P,T) = \int_{-\infty}^{+\infty} J(\lambda).d\lambda = A_{ij}\, g_i \frac{hc}{\lambda_t} \frac{1}{4\pi} \frac{n(P,T)}{Z(P,T)} e^{\frac{-E_i}{KT}} \tag{9}$$

And $\lambda_t$ is the wavelength of the light emitted by the electronic transition considered.

To summarize, for a Lorentzian profile we have:

- The area $a_l \propto J_{tot}$ and it is equal to : $a_l = A\sigma$
- $FWHM = FWHA = 2\sigma$
- $FWHA = 0.549 \left(\frac{ne}{10^{23}}\right)^{0.67965}$

The evolution of the emissivity of the H$_\alpha$ spectral line as a function of pressure and temperature, the parameters $g_i$, $A_{ij}$ and the other constants do not vary if we consider a single electronic transition. In eq. (9) $J_{tot}$ depends on pressure and temperature via the terms $n(P,T), Z(P,T)$, and $e^{\frac{-E_i}{KT}}$. For the electronic transition associated with H$_\alpha$, $E_i$ is equal to 12.0875 eV [16].

For a thermal plasma considered to be in LTE, solving the corresponding equations (Saha-Eggert, Guldberg-Waage, electrical neutrality, Dalton law) gives us the evolution of the plasma species densities as a function of pressure and temperature. The method is detailed in Harry-Solo *et al.* [12]. Thus, the species global densities and their partition function are known as a function of pressure and temperature. In our study, the composition of a pure water plasma is calculated for a pressure of 3 bar according to previous study by Z. Laforest [11]. Using these data and the eq. (9) we can determine the evolution of the emissivity of the H$_\alpha$ line as a function of temperature (see fig. 4).

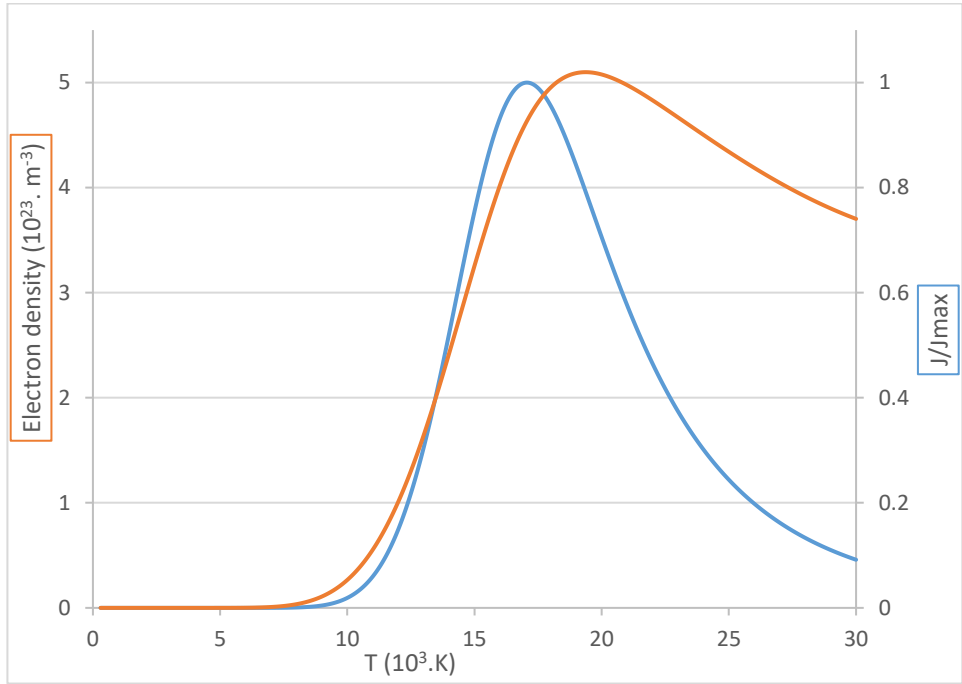

Figure 4: Evolution of the normalised emissivity of the H$_\alpha$ line (blue) and the electron density (orange) as a function of temperature for a water plasma at 3 bar.

Figure 4 shows that the maximum emissivity is reached around $17\ kK$. As indicated by eq. (9) this evolution results from the combination of different parameters. This maximum occurs at higher temperatures than the maximum density of the emissive species, hydrogen. This is explained by the contribution of the terms $Z(P,T)$ and $e^{\frac{-E_i}{KT}}$ in eq. (9) which compensate for the decrease in hydrogen density. The evolution of the emissivity was calculated using only the temperature and pressure dependent terms. The other terms are dependent on the electronic transition and have therefore been ignored as we are studying the evolution of a single transition. This is the reason why we have represented the normalised emissivity in fig. 4. It can also be noted on fig. 4 that the maximum electron density occurs at a higher temperature ($\approx 19.4\ kK$) than the maximum emissivity of the H$_\alpha$ line ($\approx 17\ kK$).

## 5    Results:
### 5.1    Different temperature profiles:

We consider an axisymmetric plasma of constant pressure (3 bar) to study the profile of the H$_\alpha$ pseudo experimental (PE) spectral line reconstructed from different temperature profiles. The composition of a pure water plasma calculated from our model [12] provides the electron density profile corresponding to the temperature profile. The objective is to study the electron density deduced from measurement of the Stark broadening of this reconstructed PE line with eq. (1) provided by Gigosos *et al.* [16]. This electron density will be compared with the electron density profile.

We arbitrarily defined four axisymmetric temperature profiles and two asymmetric ones, with a maximum temperature in the centre. These temperature profiles are common and could correspond to any type of axisymmetric plasma with the light collected in a direction perpendicular to the symmetry axis. The maximum temperature ($\approx 16\ kK$) was initially chosen to be lower than that of the maximum emissivity of the H$_\alpha$ line ($\approx 17\ kK$). These different temperature profiles are given in fig. 5 (left) with the corresponding electron density profiles (right).

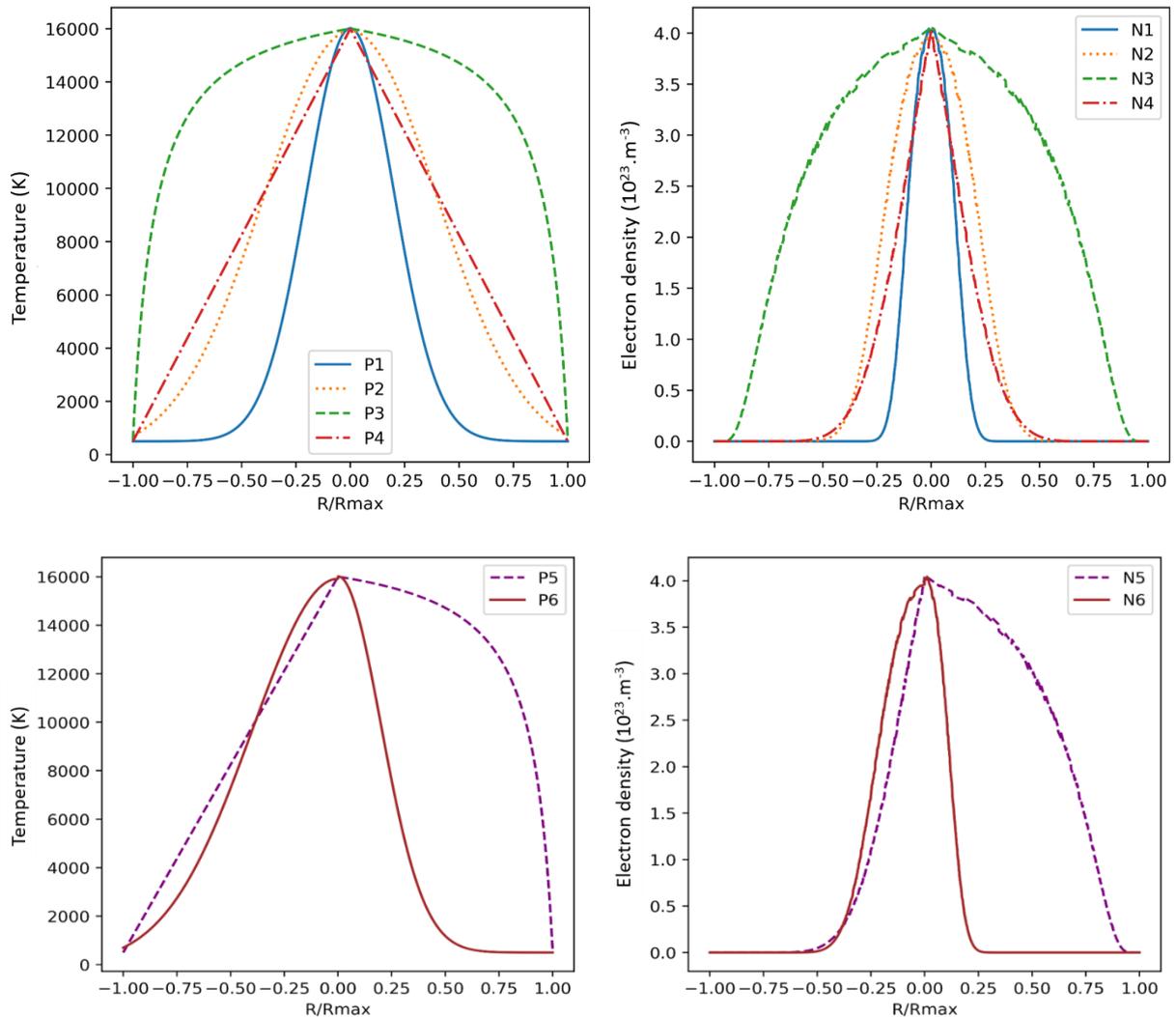

Figure 5: Temperature profiles 1 and 2 are Gaussian, profile 3 is arbitrary and profile 4 is linear. Temperature profiles 5 and 6 are asymmetric and are respectively half of profiles 3 and 4 and half of profiles 1 and 2. The electron density profiles are deduced from the plasma composition.

From these profiles and the elements determined in sect. 4, six H$_\alpha$ PE spectral lines are calculated as the sum of Lorentzian spectral lines on a plasma string whose temperature follows one of the temperature profile. The reconstruction of this spectral line was performed numerically using an in-house developed Python program. A sum of Lorentzian profiles calculated on each point of the temperature profile is performed:

$$f_{tot}(\lambda) = \sum_r f(\lambda; A, \sigma) = \frac{A}{\pi} \frac{\sigma}{\sigma^2 + \lambda^2} \qquad (10)$$

Using:

- $2\sigma = 0.549 \left(\frac{n_e(T)}{10^{23}}\right)^{0.67965}$
- $A\sigma \propto J_{tot}(T)$

Although the sum of Lorentzian profiles with different amplitudes and broadenings does not lead to a simple analytical expression of a Lorentzian, it is possible to treat this PE spectral line the same way we would an experimental one. It is common to fit integrated experimental spectral lines with the best matching Voigt or Lorentzian profile. The Lorentzian profile that best matches that of the PE spectral line is calculated using the lmfit library in Python and the broadening of this Lorentzian profile is measured.

The PE spectral line determined this way and the associated Lorentzian profile are plotted in fig. 6 for temperature profile number 4 (P4). Figure 6 shows that a Lorentzian profile allows most points to be approached correctly, with the exception of those close to the maximum. The error on the estimation of the maximum is calculated and tabulated for the different profiles. As indicated in the table of fig. 6, the discrepancy between the data and the fit is most pronounced for profile number 4. The profile number 4 (P4) is also the least realistic profile.

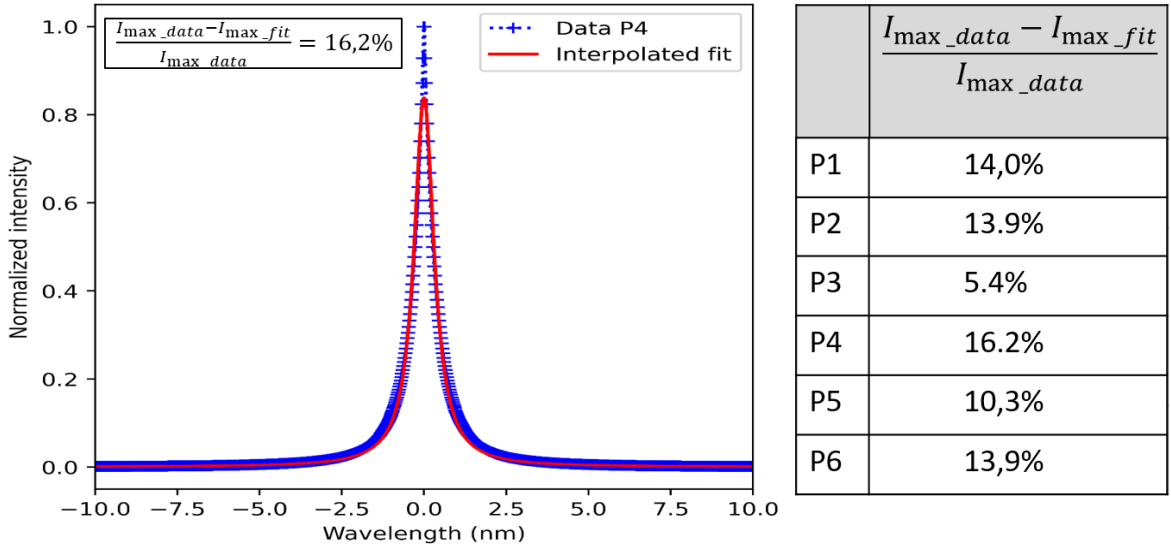

Figure 6: Comparison between the spectral line determined for the temperature profile 4 and its Lorentzian fitted profile (Left side). Table of the error in percentage on the estimation of the maximum for each profile (Right side).

We will detail the case of the fourth temperature profile (fig. 5, P4) which shows the largest deviation. We have numerically determined the FWHA of the line determined by the calculation and compared it to that of the Lorentzian profile:

$$\frac{FWHA_{data} - FWHA_{fit}}{FWHA_{data}} = 14.3\% \qquad (11)$$

We have calculated the deviation on the electron density measurement presented in the same way:

$$\frac{n_{e_{data}} - n_{e_{fit}}}{n_{e_{data}}} = 18.8\% \qquad (12)$$

This deviation is not negligible, but it has been determined for the profile with the largest deviation between the data and the fit (P4). Experimentally, we usually perform the broadening measurement (FWHA or FWHM) on the fit, and when several sources of broadening are mixed, a Voigt profile is sometimes considered with the assumption that the Lorentzian component is solely due to Stark effect. For the different lines determined, a Voigt profile does not allow a better fit of the line. However, the difference between the FWHA measurement made directly on the PE spectral line data and the fit is too large to be ignored. This is why in this study we will perform our broadening measurements on the PE spectral line data without using any fit.

The electron density profiles are calculated with eq. (1) from the broadening measurements and presented in fig. 7 (hatched in black is the mean electron density over the profile and hatched in white is the density resulting from the broadening measurement).

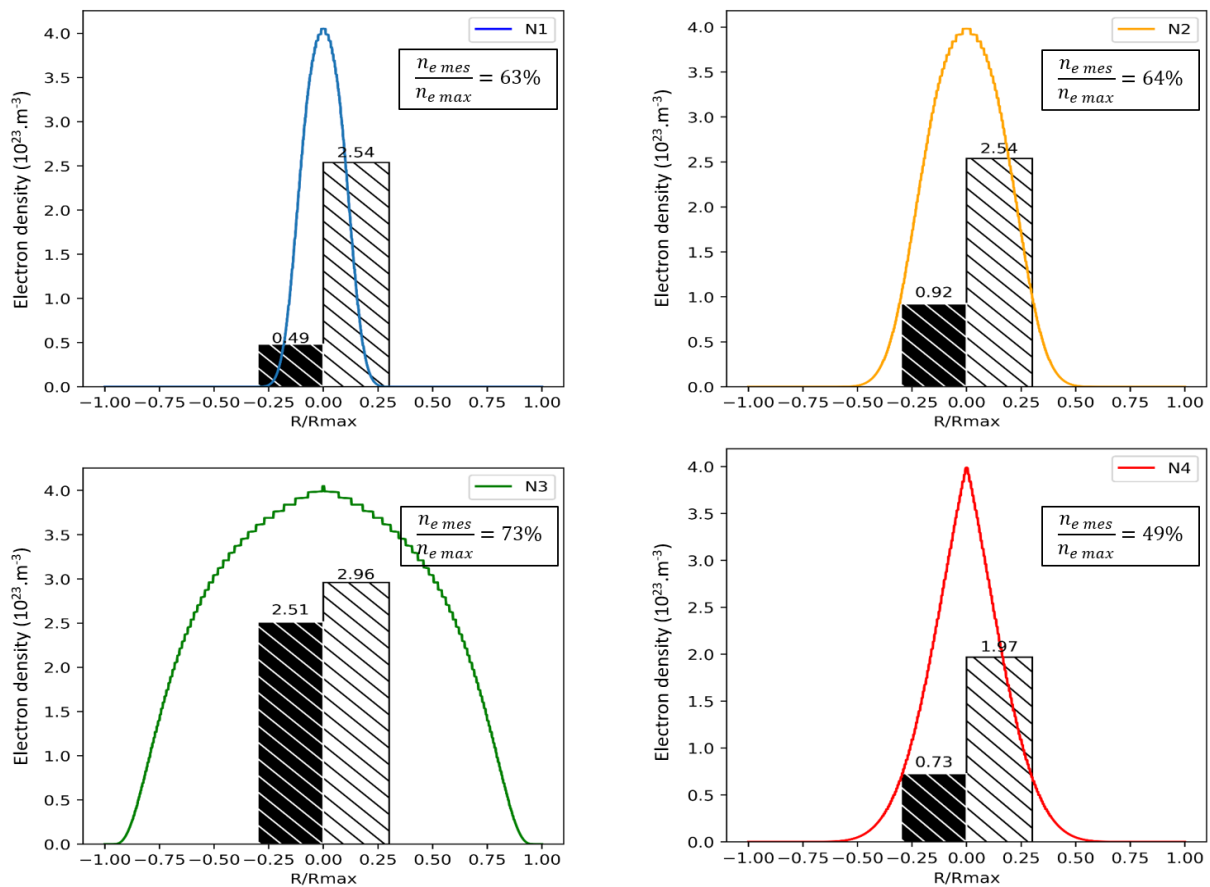

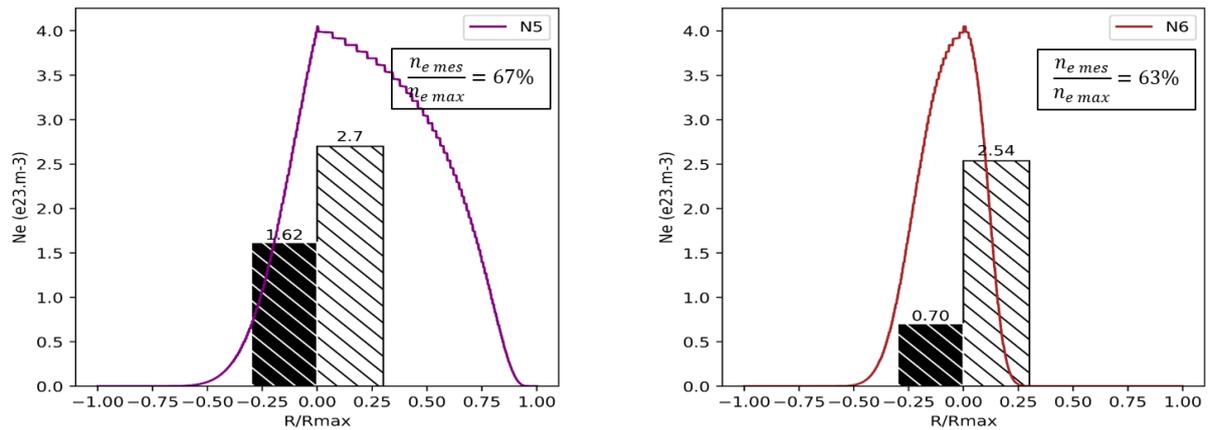

Figure 7: Electron density profiles, average over the profile and electron density measured by broadening measurement.

We observe in fig. 7 that the mean electron density for the first profile is 0.49 $10^{23}$ m$^{-3}$ and the maximum is 4.05 $10^{23}$ m$^{-3}$. The density value determined from the line broadening is 2.54 $10^{23}$ m$^{-3}$. This value is closer to the maximum than it is to the mean value of the profile. For all profiles, the density determined seems to be driven in greater part by the hottest and most electronically dense area (which will also be the most emissive area) than by the other parts of the profile. On profile P3, the electron density determined by the measurement is closest to the maximum; it is also for this profile that the warmest zone in the centre is the largest. For the temperature profile P5, which is a combination of profiles P3 and P4, the broadening measurement seems to be driven by profile P3 and less affected by the thinner (especially in the warm central zone) profile P4 which results in an electron density measurement rather close to the maximum. Profiles P1 and P2 are both Gaussian and are temperature profiles with a similar maximum temperature (and therefore with a similar maximum value of the electron density), differences exist in the peripheral zones of the plasma where profile P2 presents a higher electron densities as it is defined by a wider Gaussian profile. On the profiles P1 and P2 the densities obtained by broadening measurements are almost identical and represent the same fraction of their respective electron density profile maxima, although the average electron density is higher for profile P2. It is also true for profile P6, unsurprisingly, given that it is a combination of profiles P1 and P2. Excluding the highly unrealistic fourth profile P4, all electron densities obtained by broadening measurement represent a similar fraction of their respective electron density profile maxima, *i.e* between 63% and 73%.

It should be noted that the maximum temperature of these profiles ($\approx 16\ kK$) is lower than the maximum emissivity temperature for the H$_\alpha$ spectral line ($\approx 17\ kK$) as well as the temperature of the maximum electron density ($\approx 19.4\ kK$). In sect. 5.2 we will study the influence of the maximum temperature, especially when it exceeds these two maxima.

## 5.2   Different values of the maximum temperature:

We defined four Gaussian temperature profiles with the same width at half height but with different amplitudes. The four temperature profiles (P1-P4) and the associated electron density profiles (N1 – N4) are shown in fig. 8:

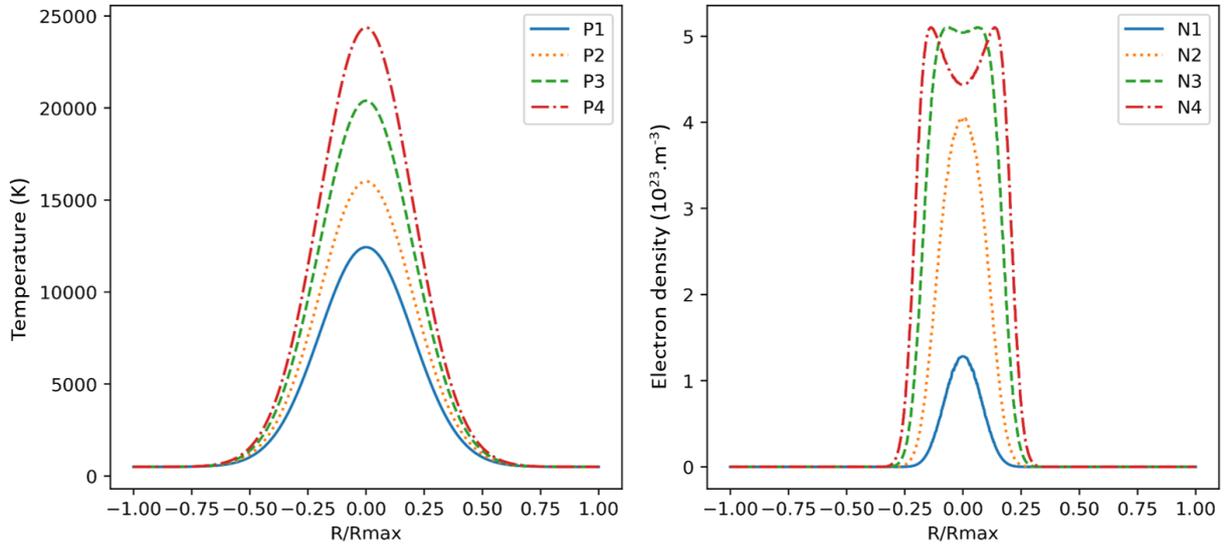

Figure 8: Gaussian temperature profiles presenting identical width at half height but different maxima. The electron density profiles are deduced from the temperature profiles.

Figure 8 shows that the maximum temperature of the first two profiles is chosen lower than the maximum emissivity temperature of the $H_\alpha$ spectral line. Profile P3 has a maximum temperature very slightly above the maximum electron density, while profile P4 has a much higher maximum temperature. We also observe in fig. 8 a dip in electron density in the centre with profile N3 which is even more pronounced on profile N4. This dip is a consequence of the temperature in the centre exceeding that of the maximum electron density. In the same way as in sect. 5.1, the measured electron density is determined from the broadening of the $H_\alpha$ spectral line calculated from these temperature profiles. These results are presented in fig. 9 with the electron density profiles (hatched in black is the mean electron density over the profile and hatched in white is the density resulting from the broadening measurement).

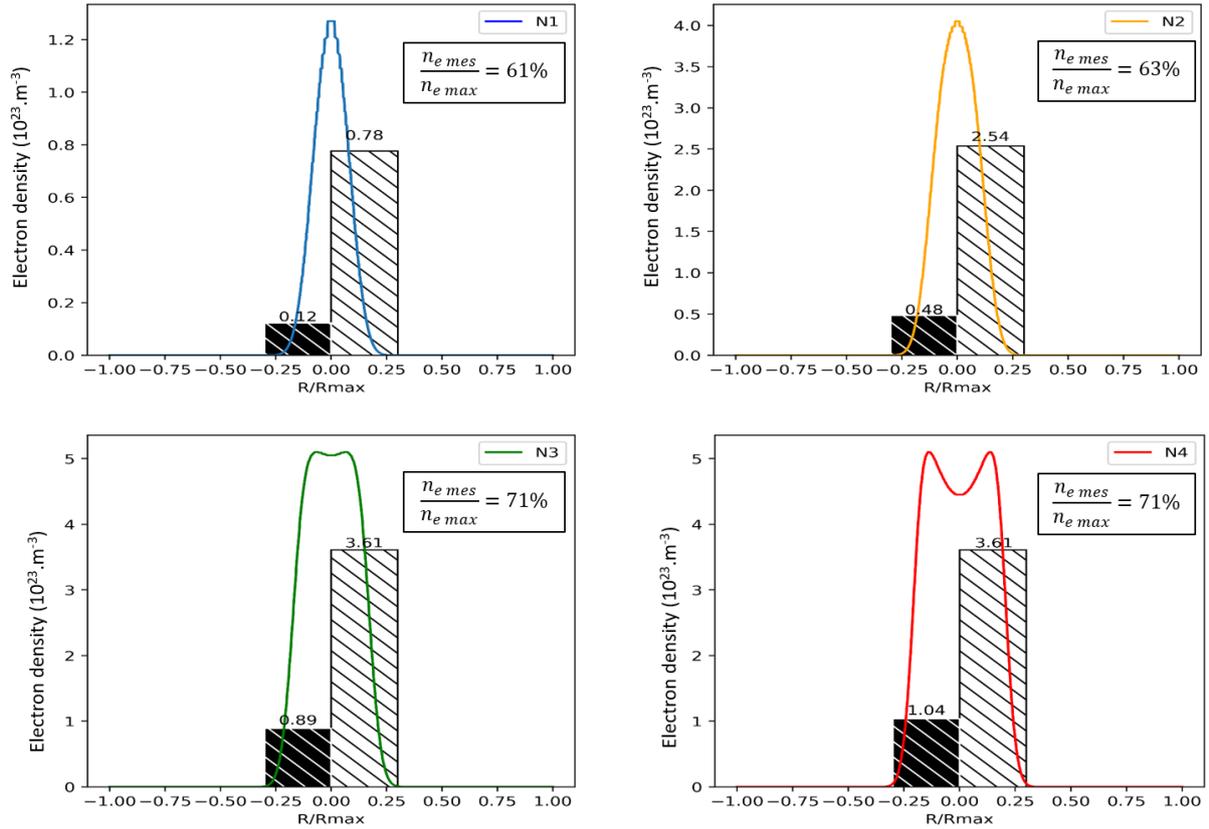

Figure 9: Electron density profiles, mean value over the profile and density obtained by broadening measurement.

The measured electron density (fig. 9) increases up to and including profile 3 (N3). Thus the maximum emissivity of the H$_\alpha$ spectral line does not seem to limit the measurement; it is still possible to measure electron densities corresponding to higher temperatures. Conversely, we do not observe an increase in electron density determined by broadening measurement on profile P4/N4 compared to that of profile P3/N3 even though the average electron density on the profile has increased. As soon as the temperature of the maximum electron density is exceeded, the electron density measured stops increasing (for a Gaussian profile maintaining the same width at half-height) which seems to indicate a similar temperature when in fact it has increased. If we consider fig. 4, we could also assume that we have exceeded the temperature of the maximum electron density (which is the case) and interpolate with higher temperatures. In this case, the temperature is overestimated. For example, we would measure more than 30 $kK$ for profile P4. Despite this, just as in sect 5.1, for all four profiles the electron density determined by broadening is somewhere between 61% and 71% of the maximum of the profile.

## 6   Conclusion:

The electron density measured from the broadening of the H$_\alpha$ PE spectral line determined as a sum over a temperature profile was studied. The study was conducted for temperature profiles with different shapes and maxima. It seems that the electron density obtained this way is governed by the maximum of the electron density profile and that it is only marginally affected by the rest of the electron density profile. The measurement is therefore rather uncorrelated with the average electron density. It was also observed that exceeding the maximum emissivity temperature of the spectral line considered did not present any particular problem. In our study, the problem lies with profiles whose temperature is such that it exceeds that of the maximum electron density at the core of the profile and induces a density dip. For these profiles, the broadening measurement is hardly usable.

It is interesting to note that the electron densities determined by broadening of the PE spectral lines represent, for all but one unrealistic profile (P4), a similar fraction of their respective electron density profiles maxima. This result is true for temperature profiles of different shapes, as shown in sect. 5.1, and for Gaussian temperature profiles with the same width at half height but with different maxima, as shown in sect. 5.2. The electron density measured by broadening measurement is 60% to 75% of the maximum electron density.

# 7   References:

1.  Vanraes P, Nikiforov A, Leys C (2012) Electrical and spectroscopic characterization of underwater plasma discharge inside rising gas bubbles. J Phys Appl Phys 45:245206. https://doi.org/10.1088/0022-3727/45/24/245206

2.  Venger R, Tmenova T, Valensi F, et al (2017) Detailed Investigation of the Electric Discharge Plasma between Copper Electrodes Immersed into Water. Atoms 5:40. https://doi.org/10.3390/atoms5040040

3.  Burakov VS, Nevar EA, Nedel'ko MI, et al (2009) Spectroscopic diagnostics for an electrical discharge plasma in a liquid. J Appl Spectrosc 76:856–863. https://doi.org/10.1007/s10812-010-9274-z

4.  Wang F, Cressault Y, Teulet P, et al (2018) Spectroscopic investigation of partial LTE assumption and plasma temperature field in pulsed MAG arcs. J Phys Appl Phys 51:255203. https://doi.org/10.1088/1361-6463/aac47f

5.  Namihira T, Sakai S, Yamaguchi T, et al (2007) Electron Temperature and Electron Density of Underwater Pulsed Discharge Plasma Produced by Solid-State Pulsed-Power Generator. IEEE Trans Plasma Sci 35:614–618. https://doi.org/10.1109/TPS.2007.896965

6.  Ni G, Zhao P, Cheng C, et al (2012) Characterization of a steam plasma jet at atmospheric pressure. Plasma Sources Sci Technol 21:015009. https://doi.org/10.1088/0963-0252/21/1/015009

7.  Mašláni A, Sember V, Hrabovský M (2017) Spectroscopic determination of temperatures in plasmas generated by arc torches. Spectrochim Acta Part B At Spectrosc 133:14–20. https://doi.org/10.1016/j.sab.2017.04.011

8. Benech J (2008) Spécificité de la mise en oeuvre de la tomographie dans le domaine de l'arc électrique : validité en imagerie médicale. These de doctorat, Toulouse 3

9. Pauna H, Aula M, Seehausen J, et al (2020) Optical Emission Spectroscopy as an Online Analysis Method in Industrial Electric Arc Furnaces. Steel Res Int 91:2000051. https://doi.org/10.1002/srin.202000051

10. Gigosos M, González M, Cardeñoso-Payo V (2003) Computer simulated Balmer-alpha, -beta and -gamma Stark line profiles for non-equilibrium plasmas diagnostics. Spectrochim Acta Part B At Spectrosc 58:1489. https://doi.org/10.1016/S0584-8547(03)00097-1

11. Laforest Z (2017) Etude expérimentale et numérique d'un arc électrique dans un liquide. These de doctorat, Toulouse 3

12. Harry Solo A, Benmouffok M, Freton P, Gonzalez J-J (2020) Stochiometry Air – $CH_4$ Mixture: Composition, Thermodynamic Propertiess and Transport Coefficients. PLASMA Phys Technol 7:21–29. https://doi.org/10.14311/ppt.2020.1.21

13. Lee WH (1980) Pressure iteration scheme for two-phase flow modeling. Multiph Transp Fundam React Saf Appl 407–432

14. Laforest Z, Gonzalez JJ, Freton P (2018) EXPERIMENTAL STUDY OF A PLASMA BUBBLE CREATED BY A WIRE EXPLOSION IN WATER. IJRRAS 34:

15. Griem H (1974) Spectral Line Broadening by Plasmas. Elsevier Science, Oxford

16. Kramida AE (2010) A critical compilation of experimental data on spectral lines and energy levels of hydrogen, deuterium, and tritium. At Data Nucl Data Tables 96:586–644. https://doi.org/10.1016/j.adt.2010.05.001